\documentclass[onecolumn,showpacs,
  superscriptaddress,%
  prd,notitlepage,showkeys,%
  10pt]{revtex4-1}

\usepackage{amssymb}
\usepackage{amsmath}
\usepackage{graphicx}
\usepackage{dcolumn}
\usepackage{fancyhdr}

\begin{document}

\title{Understanding the anomalously low dielectric constant of confined water: \protect\\an \textit{ab initio} study}

\author{T. Dufils}
\affiliation{%
Department of Physics and Astronomy, University of Manchester, Manchester, M13 9PL, UK
}
\affiliation{%
National Graphene Institute, University of Manchester, Manchester, M13 9PL, UK
}
\author{C. Schran}
\affiliation{%
Yusuf Hamied Department of Chemistry, University of Cambridge, Lensfield Road, Cambridge, CB2 1EW, UK
}
\affiliation{%
Thomas Young Centre, London Centre for Nanotechnology, and Department of Physics and Astronomy, University College London, Gower Street, London, WC1E 6BT, UK
}
\author{J. Chen}
\affiliation{%
School of Physics, Peking University, Beijing, 100871, China
}
\author{A. K. Geim}
\affiliation{%
Department of Physics and Astronomy, University of Manchester, Manchester, M13 9PL, UK
}
\affiliation{%
National Graphene Institute, University of Manchester, Manchester, M13 9PL, UK
}
\author{L. Fumagalli}
\email{laura.fumagalli@manchester.ac.uk}
\affiliation{%
Department of Physics and Astronomy, University of Manchester, Manchester, M13 9PL, UK
}
\affiliation{%
National Graphene Institute, University of Manchester, Manchester, M13 9PL, UK
}
\author{A. Michaelides}
\email{am452@cam.ac.uk}
\affiliation{%
Yusuf Hamied Department of Chemistry, University of Cambridge, Lensfield Road, Cambridge, CB2 1EW, UK
}
\affiliation{%
Thomas Young Centre, London Centre for Nanotechnology, and Department of Physics and Astronomy, University College London, Gower Street, London, WC1E 6BT, UK
}


\begin{abstract}
\hfill \break
\noindent Recent experiments have shown that the out-of-plane dielectric constant of water confined in nanoslits of graphite and hexagonal boron nitride (hBN) is vanishingly small. Despite extensive effort based mainly on classical force-field molecular dynamics (FFMD) approaches, the origin of this phenomenon is under debate. Here we used \textit{ab initio} molecular dynamics simulations (AIMD) and AIMD-trained machine learning potentials to explore the structure and electronic properties of water confined inside graphene and hBN slits. We found that the reduced dielectric constant arises mainly from the anti-parallel alignment of the water dipoles in the perpendicular direction to the surface in the first two water layers near the solid interface. Although the water molecules retain liquid-like mobility, the interfacial layers exhibit a net ferroelectric ordering and constrained hydrogen-bonding orientations which lead to much reduced polarization fluctuations in the out-of-plane direction at room temperature. Importantly, we show that this effect is independent of the distance between the two confining surfaces of the slit, and it originates in the spontaneous polarization of interfacial water. Our calculations also show no significant variations in the structure and polarization of water near graphene and hBN, despite their different electronic structures. These results are important as they offer new insight into a property of water that plays a critical role in the long-range interactions between surfaces, the electric double-layer formation, ion solvation and transport, as well as biomolecular functioning.
\end{abstract}

\maketitle

\section*{Introduction}

 \noindent The reduction of the dielectric constant of interfacial water has been the subject of extensive theoretical and experimental studies for many decades because of the ubiquitous presence of interfacial water in materials science, geology, chemistry and biology (see e.g. \cite{Stern1924DoubleLayer,Conway1951DoubleLayerDielec,Hubbard1977DielecDispersion,tielrooij2009dielectric,Netz2011DielecWater,Galli2013DielecWater,Matyushov2016,zhang_note:_2018}). However, a clear experimental demonstration of this effect proved challenging due to difficulties in probing a dielectric response of only few water layers near surfaces or in nanopores. Recently, the dielectric response of thin water layers confined in single slit-like nanochannels made of van der Waals crystals has been measured on the atomic scale \cite{fumagalli2018anomalously}. These measurements, which were conducted with nanoslits made of graphene and hBN, revealed the presence of an "electrically dead" water layer near the slit walls with a surprisingly low dielectric constant in the direction perpendicular to the surface ($\epsilon_{\perp}\approx 2 $) and extending $\sim$ 7 {\AA} into the bulk. 

The observation of an interfacial water layer with an out-of-plane dielectric constant approximately 1/40th that of bulk water sparked renewed interest in the dielectric properties of nanoconfined water. In particular, there has been a surge of theory and simulation work (see e.g. \cite{sato2018hydrophobic,Matyushov2018,%
varghese2019effect,%
ruiz2020quantifying,%
loche2020universal,%
esquivel2020anomaly,%
motevaselian2020universal,%
jalali2020out,%
Wippermann2021DielecNanoWater,%
papadopoulou2021tuning,%
munoz2021confinement,%
Bagchi2021NanoLettWater,%
ahmadabadi2021structural,%
qi2021anomalously,%
monet2021nonlocal,%
olivieri2021confined,%
cox2022dielectric}), 
including a large number of force-field molecular dynamics (FFMD) with widely used point charge models. These studies have been of tremendous value and generally predict a decrease of $\epsilon_{\perp}$ for water near surfaces, in qualitative agreement with the recent experiments. However, the intensity of the decrease, how far it extends from the surface into the bulk, and the origin of the effect are all issues under debate.
In addition, the FFMD studies involve the application of water force-fields parameterised to describe bulk rather than interfacial water properties \cite{chen2016evidence}, and the electronic coupling between water and the confining materials 
is not taken into account. Indeed, previous FFMD studies have mostly focused on water confined in graphene nanoslits. This means that it remains unclear whether the dielectric properties of nanoconfined water differ (if at all) between graphene and hBN. Such considerations call for the application of a first-principles based simulation approach such as density functional theory (DFT). DFT can deliver the requisite accuracy \cite{brandenburg2019interaction} and when combined with molecular dynamics - so-called \textit{ab initio}  molecular dynamics (AIMD) - has been used successfully to probe the structure and dynamics of nanoconfined water (see e.g. \cite{cicero2008water,bjoerneholm2016water,ruiz2020quantifying,jiang2021first}). However, DFT has yet to be used to deliver explicit estimates of the dielectric properties of confined water, mainly because of methodological challenges in extracting an accurate dielectric constant from computationally expensive AIMD simulations. 

In this study, we exploit recent developments that enable the determination of dielectric properties with AIMD \cite{sayer_finite_2019,zhang2019coupling} 
to calculate water's dielectric polarization inside graphene and hBN nanoslits. We analyze the molecular origin behind water's dielectric properties by calculating the magnitude and orientational distribution of the water molecule dipole moments as well as the topology of the hydrogen bonding network near the slit surfaces. This analysis is aided by the development of AIMD-trained machine learning (ML) potentials which ensure that converged structural properties of nanoconfined water are obtained. We show that our AIMD simulations describe the experimental results with good accuracy, predicting a much reduced $\epsilon_{\perp}$ that originates in the first two interfacial water layers, where the water dipoles are constrained in an antiparallel configuration with respect to the normal to the surface in a ferroelectric configuration. This in turn leads to a reduced effective polarization in the perpendicular direction, irrespective of the distance between the confining surfaces and their electronic structure. We contrast our DFT-based MD simulations with those obtained from classical force-field simulations revealing important differences.

\section*{Results} 
\noindent We performed simulations with AIMD, AIMD-trained ML potentials, and FFMD for water confined within graphene and hBN nanoslits, for simplicity referred to as graphene and hBN slits, as described in \emph{Materials and Methods}. To study the effect of extreme confinement on interfacial water, as in the experiments, we analyzed various separation distances between the confining layers: from 0.66 to 2.00 nm in AIMD simulations, and from 0.66 to 4.5 nm in FFMD simulations (labelled XS, S, M, L and XL, following the nomenclature used in Ref.\cite{ruiz2020quantifying}, see details in Supplementary Table S1). For each distance, we took care to establish the structure of nanoconfined water films, in agreement with previous work (see e.g. \cite{lee1984structure,cicero2008water,tocci2014friction,zhang2020water,calero2020water}) that report density oscillations in the water structure perpendicular to the surface of the confining layers. 
An example of such a density profile obtained from the AIMD-trained ML potential is shown in Fig. 1a, from which it can be seen that at least two clear solvation layers of water can be identified at each interface.

Having established the water structure, we calculated $\epsilon_{\perp}$ of water
using the finite-field method \cite{zhang_computing_2016,zhang_finite_2016,umari_ab_2002}. In short, the dielectric displacement \textbf{D} is related to the local Maxwell field \textbf{E} and the local polarization \textbf{P} by the relation $\mathbf{D}(z)=\mathbf{E}(z)+4\pi \mathbf{P}(z)$.
In the slit geometry, in the absence of free charges, the dielectric displacement in the perpendicular direction, $D_{\perp}$, is independent of $z$, therefore it is advantageous to use it here as an independent electric variable.
A variation in the local perpendicular component of the electric field, $\Delta E_{\perp}$, can then be obtained from the variation $\Delta D_{\perp}$ as
\begin{equation}
\Delta E_{\perp} = \int \epsilon^{-1}_{nl}(z,z') \Delta D_{\perp}(z')dz'
\end{equation}
\noindent where $\epsilon^{-1}_{nl}(z,z')$ is the non-local response kernel, linking the variations of $E_{\perp}$ at position $z$ to the variations of $D_{\perp}$ at position $z'$.
Because $D$ is independent of $z$, the local dielectric constant can be defined as $\epsilon^{-1}_{\perp}(z)=\int \epsilon^{-1}_{nl}(z,z') dz'$ and obtained from the local polarization response in the perpendicular direction $\Delta P_{\perp}(z)$ as
\begin{equation}
\epsilon^{-1}_{\perp}(z)=1-\dfrac{4\pi \Delta P_{\perp}(z)}{\Delta D_{\perp}}.
\end{equation}
While in previous work \cite{loche2020universal} the ratio $\dfrac{4\pi \Delta
P_{\perp}(z)}{\Delta D_{\perp}}$ was computed from the fluctuation-dissipation relation \cite{Netz2011DielecWater}, here we computed it directly from the change of polarization $\Delta P_{\perp}$ upon a change in the dielectric displacement $\Delta D_{\perp}$ using the finite-field method \cite{zhang_computing_2016,zhang_finite_2016,umari_ab_2002}, 
similarly as in Ref.\cite{zhang_note:_2018} (see \emph{Materials and Methods}).
The obtained values were corrected by subtracting the local polarization response of the system in the absence of water, that is, composed of only the confining surfaces (see Supplementary Fig. S1). This allowed us to extract the dielectric response of confined water and avoid the direct contribution of the surface, in analogy with the experimental study \cite{fumagalli2018anomalously} in which differential capacitance measurements allowed subtracting the parasitic capacitance. We then computed the dielectric constant at the level of water layers by integrating Eq.(2) over single molecular layers at various distances from the surface. To this end, we divided the water slab into water layers parallel to the surface, as indicated by the dashed vertical lines in Fig. 1b (labeled L1, L2, L3, etc.), and computed $\epsilon_{\perp}$ for each of them. The first layer, L1, has a dividing surface centered at the carbon (boron) atom position (see Supplementary Fig. S3) and is larger than the inner ones ($\sim 4.7$ {\AA} vs $\sim 3.3$ {\AA}) because it includes the depletion region between water molecules and graphene (hBN).

\begin{figure*}[h]
\centering{}
\includegraphics[width=18cm]{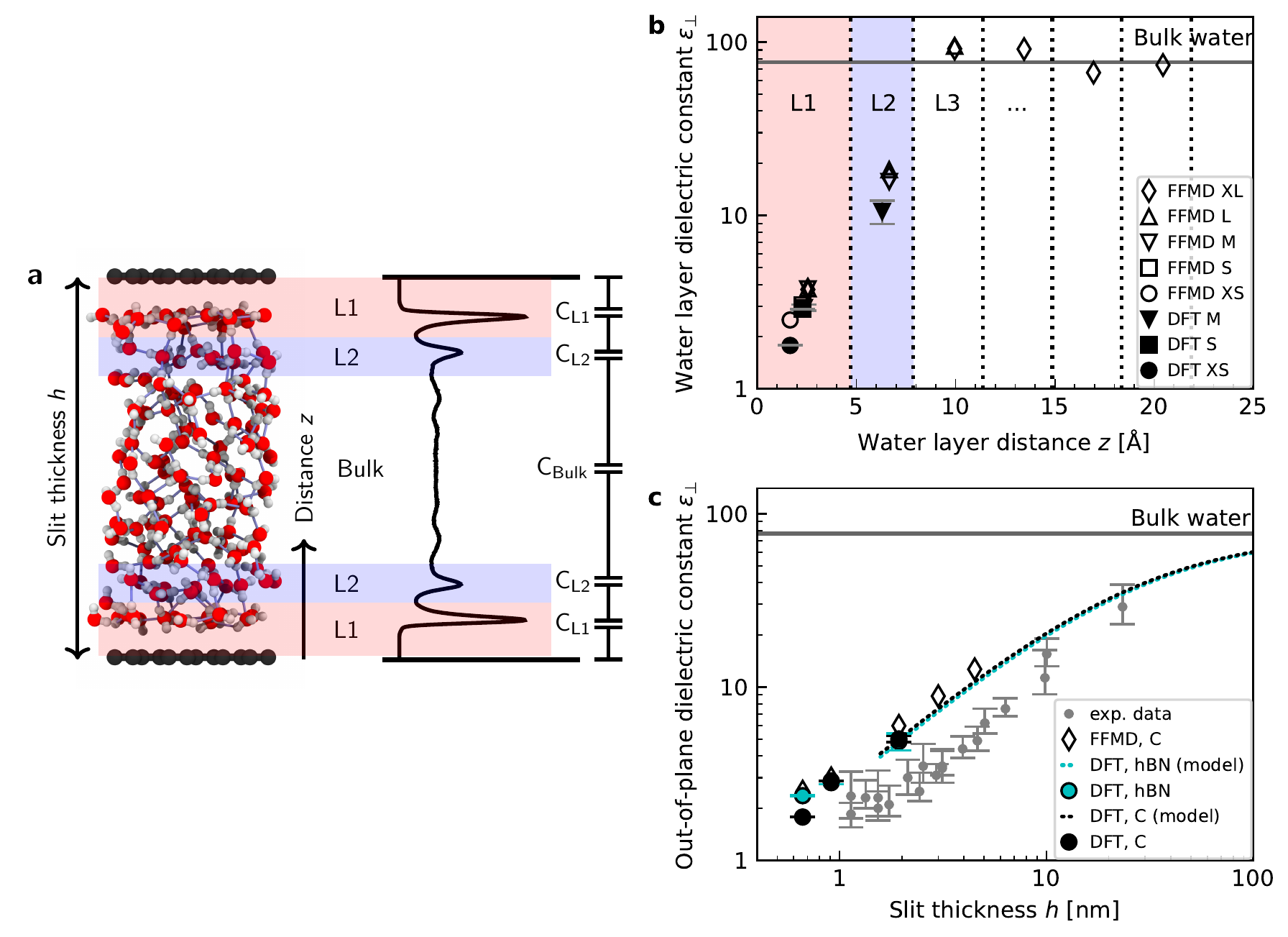}
\caption{\label{fig:diel}
\textbf{Dielectric constant of water confined between graphene and hBN.}
\textbf{(a)} Snapshot of a typical geometry for water confined within a nanoslit (left part of panel a) and the planar averaged water density profile (right part of panel a). Various terms used throughout the manuscript ($h$, $z$, $L1$, $L2$, and the capacitor model) are also indicated in this panel. 
\textbf{(b)} Calculated $\epsilon_{\perp}$ for individual water layers as a function of the distance from graphene using \textit{ab initio} (filled symbols) and force-field (open symbols) molecular dynamics calculations.
Vertical dashed lines indicate the thickness of the water layers. Light red and blue colored areas highlight the first and second interfacial layer, L1 and L2, respectively.
\textbf{(c)} Effective $\epsilon_{\perp}$ as a function of the separation distance between two confining surfaces calculated using the capacitance model in Eq.~\ref{eq:cap}.
Filled symbols are \textit{ab initio} calculations for graphene (black) and hBN (cyan) slits.
The dashed black (cyan) curve is the \textit{ab initio} prediction using the capacitor model and the interfacial $\epsilon_{\perp}$ obtained from DFT calculations for graphene (hBN) at separation distance $h= $ 1.935 nm.
Open symbols are force-field simulations for the graphene slit. 
Grey symbols are the experimental data from Ref.\cite{fumagalli2018anomalously}, shifted by 3.354 {\AA} in the x-axis direction according to the definition of $h$ in this study (atom-to-atom distance of the slit, not water thickness).}

\end{figure*}

Figure~\ref{fig:diel}b shows the calculated interfacial water-layer $\epsilon_{\perp}$ at increasing distances from graphene using both AIMD and FFMD calculations. DFT (filled symbols) yields $\epsilon_{\perp},_{\mathrm{L1}}\approx$ 2.9 for the first layer, L1 (red colored area), in good agreement with the experimental value. For the second interfacial layer, L2 (blue colored area), it yields a slightly larger value, $\epsilon_{\perp},_{\mathrm{L2}}\approx$ 11, which however remains almost one order of magnitude smaller than the bulk value.
Notably, the simulations for various thicknesses of the water slab ($h$ = 0.663, 0.911 and 1.935 nm, labelled XS, S and M, respectively) show that the computed interfacial $\epsilon_{\perp}$ for L1 and L2 are essentially independent of the slit thickness. This demonstrates that the water confinement between two surfaces is not responsible for the observed polarization suppression, rather it is an intrinsic property of interfacial water molecules at each individual interface. 

We repeated our AIMD calculations of $\epsilon_{\perp}$ for water confined within hBN slits. Despite the different electronic structure of the confining material, we found no substantial differences in the computed values as compared to the graphene slits (not shown - see Supplementary Table S2). This is to some extent expected, given that the structure of the water/graphene and water/hBN interfaces is extremely similar - see below and previous studies \cite{tocci2014friction,thiemann2022water}. Nonetheless, this is an important result, as it validates the assumption made in the experimental study, in which measurements were taken on asymmetric slits, that is, with graphite on one side of the slit and hBN on the other \cite{fumagalli2018anomalously}. 

We next turned to the force-field calculations (open symbols). The computed $\epsilon_{\perp}$ for the two interfacial layers is again in reasonable agreement with the experiments, yielding much reduced values with $\epsilon_{\perp},_{\mathrm{L1}}\approx$ 3.7 and $\epsilon_{\perp},_{\mathrm{L2}}\approx$ 16 for L1 and L2, respectively. These values are larger than the AIMD values approximately by a factor of 1.2 to 1.5, and depend little on the particular water-carbon parameterization used (see Supplementary Fig. S4). Note, however, that the FFMD values obtained here and plotted in Fig. \ref{fig:diel} do not include the contribution of the electronic polarizability. To compare them with the AIMD values, the high-frequency (electronic) dielectric constant of water, $\epsilon_{\infty}\approx$ 1.8, should be added to the calculated values, as pointed out in Ref.\cite{Matyushov2019}, and this would lead to a more pronounced deviation from the DFT values as well as from the experimental values for L1 and L2.
Importantly, our FFMD calculations show that water clearly recovers the bulk dielectric response beyond the first two interfacial layers. This reveals that water's dielectric response is insensitive to the presence of a surface at distances $z > 7.5$ {\AA}, in agreement with the experimental findings. Again, we ran simulations for various thicknesses of the water slab up to $h = 4.5$ nm and verified that these results are independent of the separation between the confining surfaces. We also ran simulations for various bulk water densities and verified that they have little effect on the predicted values (see Supplementary Fig. S5). 

Having calculated the dielectric constant of the interfacial water layers, we computed the effective dielectric constant over the whole water slab inside the slit as a function of distance between the confining surfaces, $h$, as measured in the experiment, and directly compared it with the experimental results in Fig. \ref{fig:diel}b.
To this end, we modeled the interface by three capacitors in series as in Ref.\cite{fumagalli2018anomalously}
\begin{equation}
    \dfrac{1}{C}=\dfrac{2}{C_{\mathrm{L1}}}+\dfrac{2}{C_{\mathrm{L2}}}+\dfrac{1}{C_{bulk}}
\end{equation}
where $C$ is the total capacitance, $C_{\mathrm{L1}}$ and $C_{\mathrm{L2}}$ the capacitance of the layers L1 and L2, respectively, and $C_{bulk}$ the capacitance of the remaining bulk water (see Fig. 1a), where the factor 2 arises from having two interfaces in the slit.
The dielectric constant of the water slab is thus calculated as
\begin{equation}
    \epsilon^{-1}_{\perp}(h)=\dfrac{2h_{1}}{h}\dfrac{1}{\epsilon_{\perp},_{L1}}+\dfrac{2h_{2}}{h}\dfrac{1}{\epsilon_{\perp},_{L2}}+\dfrac{h-2(h_{1}+h_{2})}{h}\dfrac{1}{\epsilon_{bulk}}
    \label{eq:cap}
\end{equation}
where $h_{1}$ = 4.72 {\AA} and $h_{2}$ = 3.15 {\AA} are the thickness of the interfacial layers L1 and L2, respectively, and $\epsilon_{\perp},_{\mathrm{L1}}$ and $\epsilon_{\perp},_{\mathrm{L2}}$ their dielectric constants previously calculated and shown in Fig.~\ref{fig:diel}a.
Figure~\ref{fig:diel}b displays the results of our calculations for both graphene (black) and hBN slits (cyan) using AIMD (filled symbols) and FFMD simulations (open symbols).
We found that our AIMD prediction describes the experimental results with good accuracy. Water's dipolar polarization in ultrathin water slabs is strongly suppressed, with $\epsilon_{\perp}\approx 2.9 $ for $h\simeq$ 1 nm, which slightly increases up to $\epsilon_{\perp}\approx 5$ for $h\simeq$ 2 nm. For thicker water slabs, an intermediate regime is observed, with $\epsilon_{\perp}$ increasing linearly with $h$ and recovering the bulk value only for 100 nm-thick water slabs, consistently with the experiment. Our force-field calculations predict a similar trend as a function of $h$, but yield slightly larger values, following from the offset found in the values of $\epsilon_{\perp}$ for the interfacial molecular layers (L1 and L2). We note that our FFMD calculations agree with recent reports in the literature \cite{zhang_note:_2018,loche2020universal,itoh2015dielectric,varghese2019effect} (also see Supplementary Fig. S4). Importantly, no significant differences were found between graphene and hBN slits, again as expected from the identical values of $\epsilon_{\perp}$ obtained for L1 and L2, as discussed above.

 \begin{figure*}[t]
\centering{}
\includegraphics[width=18cm]{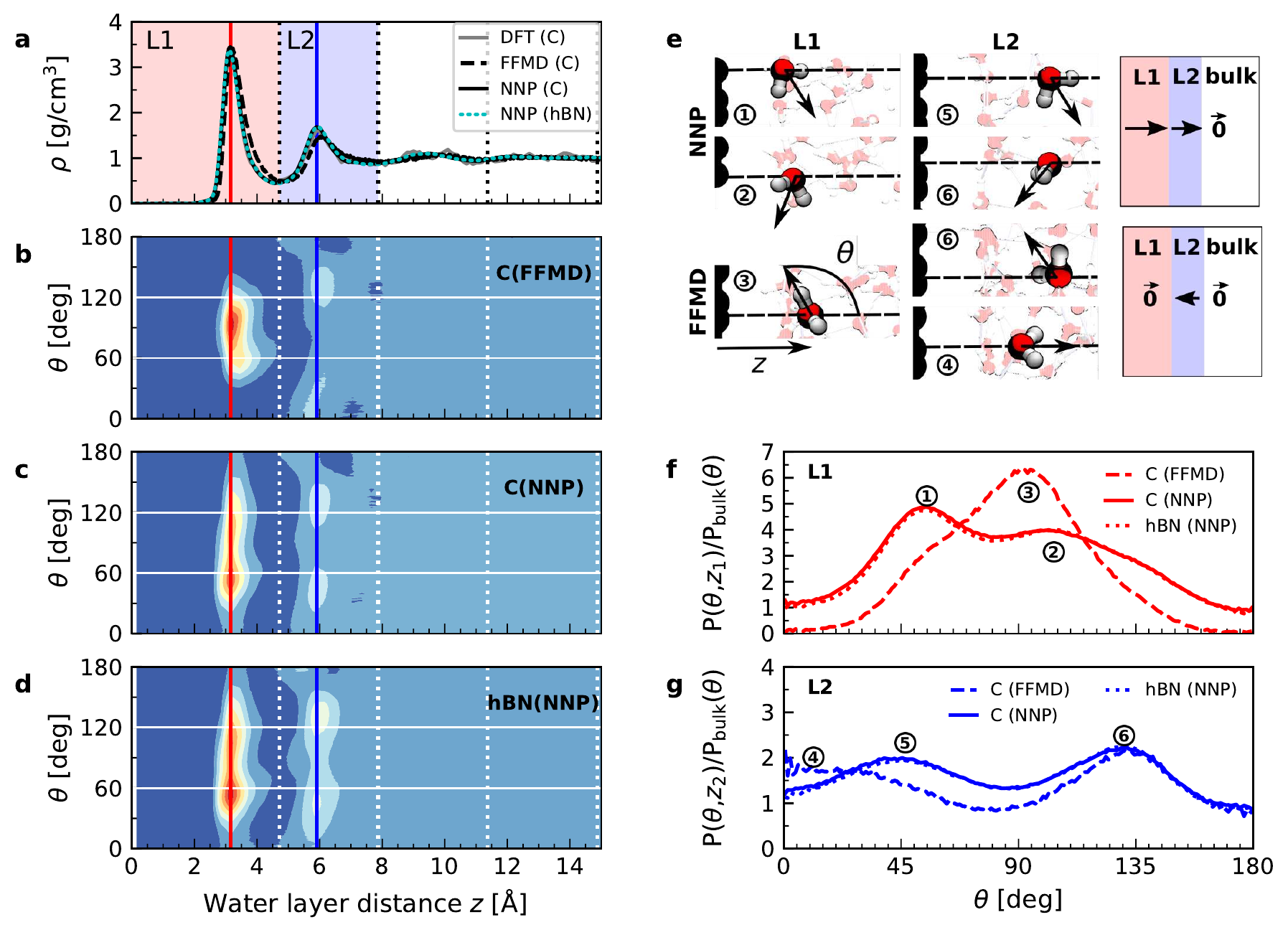}
\caption{\label{fig:water-ori}
\textbf{Orientation of water dipole moment near graphene and hBN surfaces.}
\textbf{(a)} Mass density profile obtained from AIMD (gray solid line), NNP (black solid line) and force-field (dashed black line) calculations for water on graphene and NNP calculations (cyan dashed line) for water on hBN. AIMD profiles are obtained from the average of profiles computed from $+D$ and $-D$ simulations.
\textbf{(b)} Distribution of the water dipole moment angle with respect to the perpendicular direction as a function of the distance from the surface for graphene using force-field calculations.
\textbf{(c,d)} Same as \textbf{(b)} but using NNP calculations for \textbf{(c)} graphene and \textbf{(d)} hBN.
\textbf{(e)} Snapshots of water dipole orientation in the first and second water layers, L1 and L2, at graphene (top panels) and hBN (bottom panels) surfaces using NNP calculations, showing antiparallel orientation in the perpendicular direction. The FFMD snapshot corresponding to configuration (3) shows the definition of $z$ and $\theta$. The schematics on the right represent the total net dipole moment in L1, L2 and the bulk for NNP (top) and FFMD (bottom) calculations.
\textbf{(f,g)} Dipole orientation distribution profiles at distance \textbf{(f)} $ z = 3 $ {\AA}  (L1) and \textbf{(g)} $ z = 6 $ {\AA}  (L2), corresponding to the red and blue vertical dashed lines in \textbf{(b,c,d)}, for graphene (solid line) and hBN (dotted line) using NNP calculations.
The dashed line indicates force-field calculations.
The distributions are normalized by the isotropic distribution of bulk water with density of 1 gcm$^{-3}$.}
\end{figure*}

In order to understand the origin of the observed polarization behavior of the interfacial water molecules, we analyzed the structure and the electronic properties of the nanoconfined water slabs. To this end, we obtained converged structural insight from long-time simulations using neural network potentials (NNPs)~\cite{BehlerParrinello2007,Behler2021}, which deliver DFT accuracy at zero field at substantially reduced computational cost. The NNPs for water confined by graphene and hBN were trained and validated following recent methodological developments~\cite{Schran2021} and applied for extended simulations at various confinement widths (see \emph{Materials and Methods}).
This enabled us to calculate in detail the time-averaged structural arrangement of water molecules  and hydrogen-bonding network as a function of the distance from the confining surfaces. Figure~\ref{fig:water-ori} shows some of these results at both graphene and hBN interfaces.
As expected, we found no differences in the mass density profiles of water near graphene and hBN (Fig. 2a), which show two density maxima corresponding to the two interfacial layers (L1 and L2).
Notably, the time-averaged orientational distributions obtained here reveal that water also has very similar orientations near graphene (Fig.~\ref{fig:water-ori}c) and hBN (Fig.~\ref{fig:water-ori}d). 
Indeed, there is an almost quantitative match in L1 and L2, as verified by plotting the distribution profile at the first two density maxima (Fig.~\ref{fig:water-ori}f and Fig.~\ref{fig:water-ori}g, respectively).
Furthermore, they indicate that the water molecules are strongly oriented in two preferential directions in the first two interfacial layers. %
Taking $\theta$ as the angle between the surface normal and the water molecule dipole (see Fig.~\ref{fig:water-ori}e), we find that 
in the first layer, the water dipole moments are oriented either towards the bulk water at $\theta\approx 60^{\circ}$ or towards the surface at $\theta\approx 105^{\circ}$.
These two preferential orientations are also found in the second layer,
with a peak at $\theta\approx 45^{\circ}$ and the other at $\theta\approx  135^{\circ}$.
Thus, the molecules in the first two interfacial layers are divided into two distinct  and  anti-parallel populations with respect to the dipole projection on the surface normal: either leaning towards the surface or towards the bulk. The effective dipolar polarization in the perpendicular direction that could be derived from the NNP simulations is greatly reduced compared to the water molecular dipole, in agreement with the much reduced dielectric response that was obtained here and in the experiment. These observations confirm the recent suggestion of antiparallel perpendicular components of interfacial dipoles \cite{Review2021Matyushov} and the proposed Ising model, where the perpendicular dipole moment can only take two values, positive and negative \cite{Bagchi2021IsingModel}, at the origin of water's polarization suppression near the surface. However, both these two models did not capture the complexity of the water structure observed here with our detailed analysis. 
We also note that the force-field simulations tell a rather different story: consistent with previous force-field simulations \cite{Rossky1984WaterStructure,varghese2019effect,ahmadabadi2021structural,olivieri2021confined}, the dipoles in the first layer (Fig.~\ref{fig:water-ori}b) are primarily oriented in a single dominant direction parallel to the surface. The second layer is more bimodal in nature but again different from the DFT distribution through the presence of molecules aligned precisely with the surface normal. 

The dipole distribution profiles shown in Fig.~\ref{fig:water-ori}f and Fig.~\ref{fig:water-ori}g clearly show a breaking of symmetry in the two dipole orientations yielded by our NNP calculations. This translates into a residual net dipole moment in the perpendicular direction directed towards the bulk water in both the first and second interfacial layers, as sketched in Fig.~\ref{fig:water-ori}e. This net polarization persists in the absence of an externally applied field, therefore it is ferroelectric in character and generates a local Maxwell field. We estimate the average field generated by the spontaneous polarization of the interfacial water to be around 6.5 Vnm$^{-1}$ in the first layer and 3.0 Vnm$^{-1}$ in the second layer using the relation $\mathbf{E}(z)=-4\pi \mathbf{P}(z)$ valid in the absence of an external electric displacement. We note that the force-field calculations also predict a ferroelectric order of interfacial water, but different from the one obtained by DFT calculations, also sketched in Fig.~\ref{fig:water-ori}e.
They yield a total net dipole only in the second layer oriented towards the surface, and not the bulk, while a net zero dipole is obtained in the first layer with the water dipoles oriented parallel to the surface. 

\begin{figure}[b]
\centering{}
\includegraphics[width=9cm]{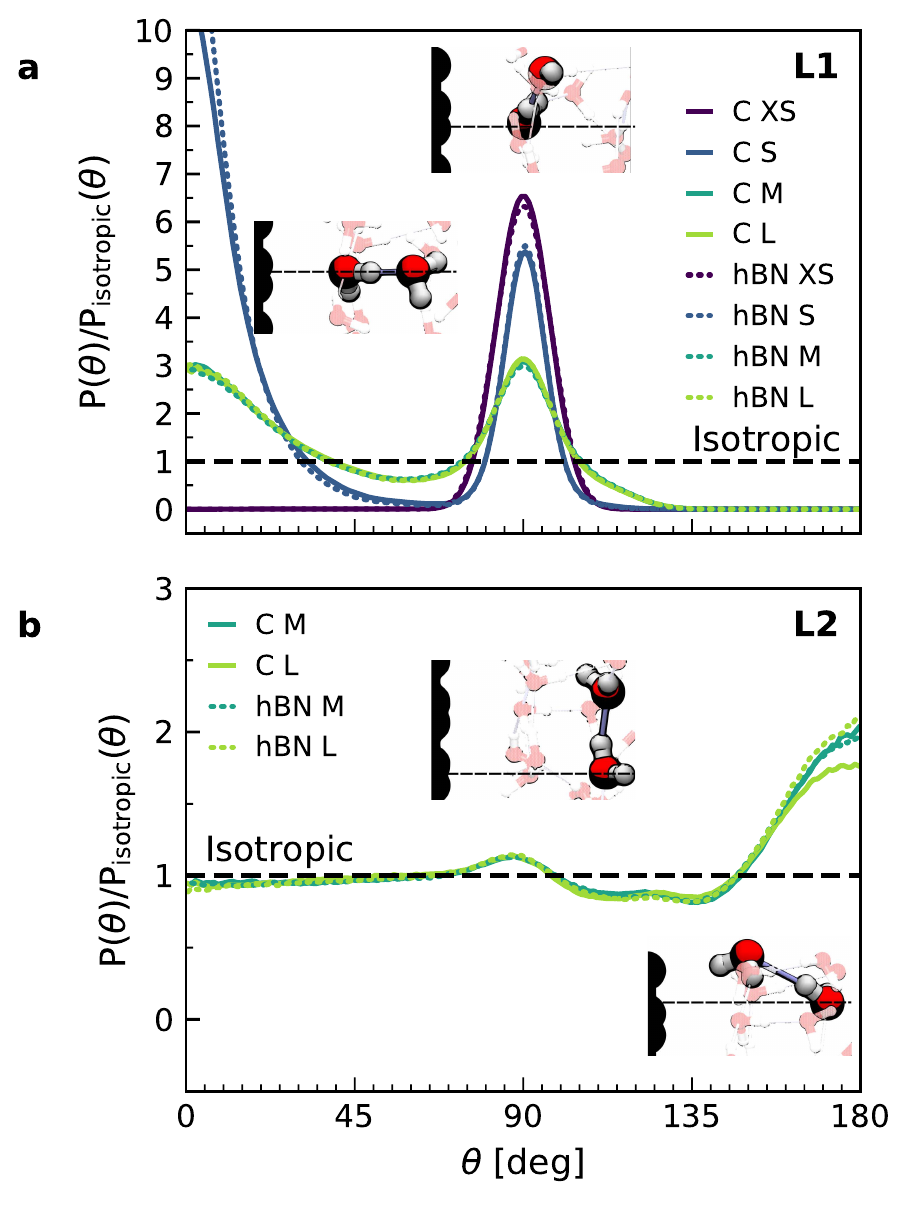}
\caption{\label{fig:hbond-ori}
\textbf{Orientation of hydrogen bonds in the interfacial water layers.}
Calculated orientation distribution with respect to the normal direction of the hydrogen bonds for water on graphene (solid lines) and hBN (dashed lines) in the first \textbf{(a)} and second \textbf{(b)} interfacial layers, normalized by the isotropic distribution, obtained from the combined NNP and DFT calculations (see FFMD simulations in Supplementary Fig. S6).
The direction corresponds to the donor oxygen-acceptor oxygen.
Inset: snapshots of the main hydrogen bonding orientations.}  %
\end{figure}

Let us now analyze the nature of the hydrogen bonding network. Figure~\ref{fig:hbond-ori} shows the orientation distribution of the hydrogen bonds, which we calculated from NNP simulations using a common hydrogen bond criterion \cite{Hbond_criteria} (see details in SI).
In the first layer, the hydrogen bond orientations are split into two populations: in-plane ($\theta\approx 90^{\circ}$) and out-of-plane pointing towards the second layer ($\theta\approx 0^{\circ}$). As expected, the latter disappears in the case of a water monolayer confined inside the slit (Fig.~\ref{fig:hbond-ori}a, blue line), as the out-of-plane hydrogen bonds are absent. Similar orientations are also observed in force-field calculations (see Supplementary Fig. S6). However, our NNP simulations yield a larger proportion of out-of-plane hydrogen bonds and in a smaller angular window, indicating a more constrained hydrogen-bonding network.
The second layer displays an almost bulk-like behavior, with only a slight increase in in-plane hydrogen bonds, related to the dipole orientation at $\theta\approx 130^{\circ}$, and in out-of-plane hydrogen bonds pointing towards the first layer.  

\begin{figure}[t!]
\centering{}
\includegraphics[width=9cm]{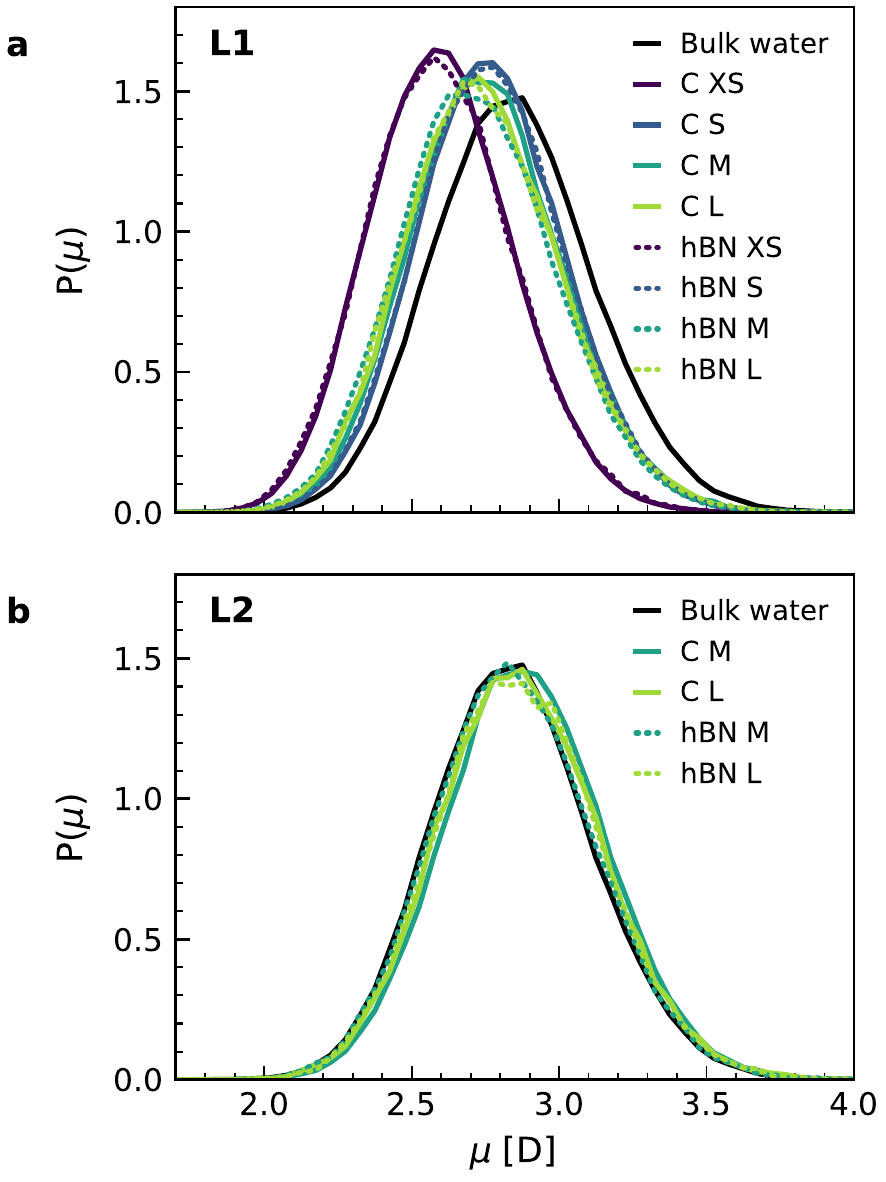}
\caption{\label{fig:dipole}
\textbf{Dipole moment distribution in the interfacial water layers.}
Dipole moment distributions obtained from the combined NNP and DFT calculations for the first \textbf{(a)} and second \textbf{(b)} interfacial layers of water on graphene (solid lines) and hBN (dashed lines), as well as bulk water. The distribution is normalized by the isotropic distribution}
\end{figure}

Finally, we calculated the molecular dipole moment in the first two interfacial layers using a combination of DFT and NNPs (Fig.~\ref{fig:dipole}) as it may also change with the change in the hydrogen-bonding network near the surface. Our results show a reduction in the dipole moment of the water molecules, but only in the first interfacial layer (Fig.~\ref{fig:dipole}a), not in the second one (Fig.~\ref{fig:dipole}b), with no appreciable dependence on the orientation of the water molecules.
The decrease in the first layer (around 4\%) is related to the decrease of the average number of hydrogen bonds per molecule in the first layer (2.93) as compared to the bulk (3.46), while this number remains bulk-like in the second layer (3.49).
Notably, for the water monolayer confined inside the slit, it is more pronounced (around 9\%) and is associated with a further decrease in the number of hydrogen bonds per molecule down to 2.18.
The observed decrease in the water molecular dipole contributes to the reduction of the dipolar polarization response and is not captured in the most commonly used force-field models, given the fixed nature of the water dipole moment.
Although being a relatively minor contribution, this effect further accentuates the difference between first-principles and force-field interfacial water.
As with the other properties, and despite the different electronic structures of the confining materials, we found no significant differences in the dipole moment distributions for water confined inside graphene and hBN slits. 

\section*{Discussion} 

\noindent In this study, we performed AIMD, AIMD-trained NNP simulations, and FFMD simulations of water confined within graphene and hBN slits aimed at understanding the out-of-plane dielectric constant of confined water.
Our AIMD calculations reproduce the experimental results of Ref.\cite{fumagalli2018anomalously} with good accuracy, yielding an ``electrically dead'' interfacial layer of $\sim$ 7 {\AA} thickness. This corresponds to a much reduced $\epsilon_{\perp}$ in the first two interfacial water layers, an effect that is seen for both graphene and hBN interfaces and is independent of the distance between the confining surfaces of the slit. 
Our AIMD and NNP MD simulations reveal that this arises from two dominant effects: the spontaneous antiparallel alignment of the out-of-plane component of water dipoles, leading to a residual ferroelectric-like polarization of these two water layers, and the relatively constrained hydrogen bonds in the first two layers. Our simulations also reveal a small reduction in the dipole moments of the individual water molecules in the first layer. Taken together, these effects are responsible for the suppression of dipolar and hydrogen bond fluctuations in the out-of-plane direction, resulting in the anomalously low dielectric constant of confined water.

Force-field simulations performed as part of this study also reveal a reduced $\epsilon_{\perp}$ of confined water, in line with several similar studies \cite{zhang_note:_2018,loche2020universal,itoh2015dielectric,varghese2019effect}. However, the extent of the reduction is underestimated in the force-field results compared to the DFT. Partly this is down to the different interfacial water structures predicted by DFT and the force-fields, and partly because the force-field models fail to capture the small reduction in dipole moments of the interfacial (L1) water molecules. Although DFT is not without its own shortcomings, the functional used here has been validated against high-level electronic structure theory (Quantum Monte Carlo, see \textit{Materials and Methods}). Specifically, it accurately describes the relative energy of different two-dimensional (2D) ice structures and the water-carbon interaction strength. Standard point charge force-field models, while being incredibly useful in deepening understanding of water in general, have been designed for bulk water and cannot be expected to deliver quantitative accuracy for interfacial water. 

The emergence of ferroelectricity in interfacial and confined water due to the spontaneous alignment of water dipoles near surfaces has been under debate for many years. In particular, water has been predicted to arrange in ferroelectric or antiferroelectric configurations when confined in one-dimensional channels such as nanotubes and nanopores \cite{Dellago2008,FerroelecWater2011} and in 2D nanoslits in the in-plane direction \cite{InPlaneFerroelecWater2016}, while experimental evidences remain scarce and are generally associated to ice phases \cite{FerroelecIceSamorjai1998,FerroelecIceIedema1998,Bramwell1999}.
Our first-principles calculations provide insight into this question as they indicate that while retaining liquid-like mobility, interfacial water near hydrophobic surfaces is ferroelectric in the out-of-plane direction at room temperature, irrespective of the presence of two confining surfaces. Importantly, the "electrically dead" water layers reported in Ref.\cite{fumagalli2018anomalously} should arise from this effect. Note that when water is under strong confinement between two hydrophobic surfaces, as in the experiment, the interfacial water at the two confining surfaces exhibits a net polarization towards the bulk water in the center of the slit (see sketch in Fig. 2e), thus ordering in opposite directions with respect to the surface normal in an overall antiferroelectric configuration in the slit. From the calculated net polarization, we estimate the coercive external field required to reorient it to be at least a few Vnm$^{-1}$, which is much larger than the electric field which was effectively applied to the water in the experiment, estimated to be around 0.05 Vnm$^{-1}$ \cite{fumagalli2018anomalously}. This would explain why the interfacial water dipoles remained aligned in their spontaneous configuration near the surfaces in the experiment and did not reorient in the opposite direction with the external field. 
Given the general nature of confinement in our study and that much of surface chemistry and molecular biology is influenced by the dielectric properties of water at interfaces, we expect these findings to be of broader relevance, in particular for understanding water's polarization near and across biological membranes and for developing new energy storage/conversion devices.

\section*{Materials and methods}
\textbf{\textit{Ab initio} molecular dynamics calculations.}  
AIMD simulations were performed with the CP2K \cite{doi:10.1063/5.0007045} code. 
We simulated boxes with a lateral size of 1.23$\times$1.28 nm corresponding to 5x3 supercells of the orthorhombic unit cells of the two-dimensional confinement materials.
These comprise 60 carbon atoms or 30 BN pairs per layer of graphene and hBN, respectively.
We used the same lateral dimensions of the cell for both graphene and hBN due to their very similar lattice constants \cite{graziano2012improved} (see details in Supplementary Table S1). 
The separation between the two confining surfaces, $ h $, varied from 0.663 (XS), to 0.911 (S), to 1.935 (M) nm. 
2.0 nm of vacuum was introduced to prevent interactions between periodic images in the normal direction. 
We used the revPBE functional \cite{zhang1998comment} with the D3 \cite{grimme2010consistent} correction for dispersion interactions, given its good performance in accurately reproducing both the experimentally measured structure and dynamics of liquid water \cite{Morawietz2016,gillan_perspective:_2016,Marsalek2017} as well as the interaction energies of water on graphene and inside CNTs obtained using more advanced methods such as Quantum Monte Carlo and coupled cluster theory \cite{brandenburg2019interaction}.
Goedecker-Teter-Hutter (GTH) pseudopotentials \cite{goedecker1996separable} 
were used along with a DZVP basis set for carbon, boron, and nitrogen and a TZV2P basis set for oxygen and hydrogen in conjunction with a 750 Ry plane-wave cutoff. %
Simulations were performed in the  NVT ensemble with a target temperature of 330 K using a Langevin thermostat, a timestep of 1.0 fs, and deuterium masses for the hydrogen atoms. 
The surface atoms were kept fixed and initial configurations were generated using the fftool and packmol packages \cite{martinez2009packmol}. 
These simulation cells were first equilibrated using FFMD calculations using the force field and simulation parameters described below.
The number of water molecules was adjusted such that the mass density profiles match the ones obtained for the larger simulation cells used for the FFMD simulations described below.
Each simulation was first equilibrated for 3 ps, while all shown properties were accumulated over at least 60 ps additional simulation time (see details in the SI and Supplementary Fig. S2).
The dipole moment distribution of water was calculated with DFT using the same functional and set-up as described above.
For this purpose we computed the position of the Wannier centers of the water molecules using uncorrelated structures obtained from the NNP trajectory (see below).
Values of the dipole moments reported in Fig. 4 were obtained by averaging data obtained every 0.8 ps from the first 5 ns (XS and S), 3 ns (M) and 2 ns (L) NNP trajectories described below.

\textbf{Force-field molecular dynamics calculations.} 
We simulated boxes of lateral size of 3.76 nm with two confining layers at distance $ h $ 0.663 (XS), 0.911 (S), 1.935 (M), 3.00 (L) and 4.50 nm (XL) using the same lattice constant as in the AIMD simulations. 
Simulations were performed in the NVT ensemble at 300 K using a Nos\'e-Hoover thermostat with a time constant of 1 ps and a timestep of 2 fs. 
Each simulation was equilibrated for 2 ns, while data were accumulated for the following 5 ns.
For simulations without an applied field, we used the LAMMPS package \cite{plimpton1995fast}, with initial configurations generated using the fftool and packmol packages \cite{martinez2009packmol}. 
The number of water molecules was adjusted such that the density was 1 g/cm$^{3}$.
The density was defined as the mass of water molecules divided by the accessible volume, where the accessible volume corresponds to the region in space where the atomic number density (either oxygen or hydrogen) is non zero.
For the simulations in the constant-D ensemble, we used the Metalwalls software \cite{marin2020metalwalls}. 
The water molecules were modeled with the SPCE model \cite{berendsen1987missing} and three different water-carbon interaction potentials \cite{steele1974interaction,ruiz2020quantifying,brandenburg2019physisorption}.
All data shown in the main text were obtained with the water-carbon parameterization from \cite{ruiz2020quantifying}. Data using other water-carbon models are provided in the SI. 
For the water-hBN interaction, we used the force-field of \cite{wu_hexagonal_2016}.

\textbf{Dielectric constant calculations.} 
The perpendicular dielectric constant was computed following the formalism of Ref.\cite{Netz2011DielecWater} and using the finite-field method, applying an electric displacement of $\pm$ 1.0 Vnm$^{-1}$ for DFT calculations and $\pm$ 1.5 Vnm$^{-1}$ for FFMD calculations. 
The lower electric displacement value used in the DFT calculations was chosen to be safely within the linear regime, while this is estimated to be 1.5-2.0 Vnm$^{-1}$ for force-field calculations. 
To improve convergence, we computed the difference in polarization between positive and negative dielectric displacement, $+D$ and $-D$, effectively doubling the window of the linear regime.
The dielectric displacement is applied using the same procedure as in \cite{zhang2019coupling}, implying 3D periodic boundary conditions, along with Ewald summation for the Coulomb interactions. 
For the AIMD calculations, we computed the polarization from the full charge density (electrons plus nuclei), thus taking into account the multiple dipole orders in the polarization derivation. 
To remove the contribution of the confining surfaces, we carried out a differential calculation, in which we subtracted the response of the confining surfaces without water to the local response of the total system (see details in SI and Supplementary Fig. S1). As the atoms of the surface were kept fixed, only a single calculation was required to compute the former. 
In the case of AIMD calculations, the convergence of the dielectric profile proved to be slow (not converging after 70 ps of simulations). 
On the other hand, the dielectric constant converged when integrating the dielectric profile over a molecular layer, as discussed in details in SI (Supplementary Fig. S2).
We note that the scale of the molecular layer (a slice with a width of around 0.4 nm and an infinite extension along the interface directions) is more relevant for this study, as Eqs.(1-2) have been derived in the scope of continuous media theory. 
Thus, we extracted the dielectric response of the first two interfacial layers, which tend to converge faster due to their small value, and the constant-D ensemble, which allowed us to directly obtain the inverse of the dielectric constant. 
To limit the computational cost of AIMD calculations, we assumed bulk-like dielectric response for the remaining water molecules in the center of the slit, as found in the force-field calculations.

\textbf{Machine learning potentials.}
In order to access converged structural properties of water confined by graphene and hBN layers, we made use of machine learning potentials to perform long-time simulations at DFT accuracy.
For that purpose, we utilized Behler-Parrinello neural network potentials (NNPs)~\cite{BehlerParrinello2007,Behler2021} in a committee model~\cite{Schran2020} enabling the simple development~\cite{Schran2021} of machine learning potentials at first-principle accuracy for the two confining materials.
Using this methodology, we trained and validated the two models as described in detail in Ref.~\citenum{thiemann2022water} and applied them here for MD simulations at various surface separations, always comparing graphene and hBN confinement.
As before, simulations were performed with CP2K in the NVT ensemble at 330\,K using a Langevin thermostat, a timestep of 1.0\,fs, deuterium masses for the hydrogen atoms and keeping the atoms of the confinement material fixed.
Systems with two confining layers at distance $ h= $ 0.663 (XS), 0.911 (S), 1.935 (M), 3.00 (L) and a lateral size of 1.23$\times$1.28 nm were propagated for 5\,ns to converge structural properties.

\section*{Authors contribution}
\noindent L.F., T.D., C.S. and A.M. designed the research; T.D. performed the simulations and C.S. developed the machine learning models; all authors analyzed the data; and L.F., T.D., C.S., and A.M. wrote the paper, with contributions from J.C. and A.K.G.

\begin{acknowledgments}
The authors thank D. V. Matyushov for fruitful discussions.
T.D. and L.F. were supported by the European Research Council (grant 819417) under the European Union Horizon 2020 research and innovation programme.
C.S. acknowledges partial financial support from the \textit{Alexander von Humboldt-Stiftung}.
J.C. and A.M. thank Beijing Natural Science Foundation (No. JQ22001) for support.
We are grateful to the UK Materials and Molecular Modelling Hub for computational resources, which is partially funded by EPSRC (EP/P020194/1 and EP/T022213/1),
and computational support from the UK national high performance computing service, ARCHER, for which access was obtained via the UKCP consortium, funded by EPSRC grant ref EP/P022561/1. 
We are also grateful to the Research IT and the use of the High Performance Computing (HPC) Pool funded by the Research Lifecycle Programme at the University of Manchester. 
\end{acknowledgments}

\hfill \break
\hfill \break
%
%

\end{document}


\maketitle

\tableofcontents
\clearpage

\section{Simulation details}
All relevant information about the simulations performed in this study are given in Table~\ref{tab:sim-det} below.

\begin{table}[h!]
\centering{}
\caption{\textbf{Details of the AIMD and FFMD simulations}.
For each slit thickness $h$, defined as the separation between the two confining materials, we provide the cell dimensions $L_{x/y/z}$, the number of atoms $N_\text{atoms}$ and water molecules $N_\text{water}$, the equilibration period $t_\text{eq}$ and the total simulation length $t_\text{sim}$.
\label{tab:sim-det}}
\begin{tabular}{p{3cm}p{4cm}p{4cm}p{4cm}}
\hline
\multirow{2}{3cm}{Slit thickness} & \multicolumn{3}{c}{Simulation details} \\
 & Graphene (AIMD) & hBN (AIMD) & Graphene (FFMD)\\
 \hline
\multirow{7}{3cm}{XS \\ $h$ = 6.63\,\AA} & $L_x$ = 12.30\,\AA & $L_x$ = 12.30\,\AA & $L_x$ = 36.89\,\AA\\
 & $L_y$ = 12.78\,\AA & $L_y$ = 12.78\,\AA & $L_y$ = 38.34\,\AA\\
 & $L_z$ = 26.63\,\AA & $L_z$ = 26.63\,\AA & $L_z$ = 80.0\,\AA \\
 & $N_\text{atoms}$ = 120 & $N_\text{atoms}$ = 120 & $N_\text{atoms}$ = 1080\\
 & $N_\text{water}$ = 16& $N_\text{water}$ = 16& $N_\text{water}$ = 147\\
 & $t_\text{eq}$ = 3\,ps& $t_\text{eq}$ = 2\,ps& $t_\text{eq}$ = 2\,ns\\
 & $t_\text{sim}$ = 100\,ps& $t_\text{sim}$ = 100\,ps& $t_\text{sim}$ = 5\,ns\\
 \hline

\multirow{7}{3cm}{S \\ $h$ = 9.11\,\AA} & $L_x$ = 12.30\,\AA & $L_x$ = 12.30\,\AA & $L_x$ = 36.89\,\AA\\
 & $L_y$ = 12.78\,\AA & $L_y$ = 12.78\,\AA & $L_y$ = 38.34\,\AA\\
 & $L_z$ = 29.11\,\AA & $L_z$ = 29.11\,\AA & $L_z$ = 80.0\,\AA \\
 & $N_\text{atoms}$ = 120 & $N_\text{atoms}$ = 120 & $N_\text{atoms}$ = 1080\\
 & $N_\text{water}$ =  30& $N_\text{water}$ = 30& $N_\text{water}$ = 268\\
 & $t_\text{eq}$ = 2.5\,ps& $t_\text{eq}$ = 2\,ps& $t_\text{eq}$ = 2\,ns\\
 & $t_\text{sim}$ = 80\,ps& $t_\text{sim}$ = 80\,ps& $t_\text{sim}$ = 5\,ns\\
 \hline
 
\multirow{7}{3cm}{M \\ $h$ = 19.35\,\AA} & $L_x$ = 12.30\,\AA & $L_x$ = 12.30\,\AA & $L_x$ = 36.89\,\AA\\
 & $L_y$ = 12.78\,\AA & $L_y$ = 12.78\,\AA & $L_y$ = 38.34\,\AA\\
 & $L_z$ = 39.35\,\AA & $L_z$ = 39.35\,\AA & $L_z$ = 80.0\,\AA \\
 & $N_\text{atoms}$ = 120 & $N_\text{atoms}$ = 120 & $N_\text{atoms}$ = 1080\\
 & $N_\text{water}$ = 82 & $N_\text{water}$ = 82 & $N_\text{water}$ = 754\\
 & $t_\text{eq}$ = 4\,ps& $t_\text{eq}$ = 2\,ps& $t_\text{eq}$ = 2\,ns\\
 & $t_\text{sim}$ = 120\,ps& $t_\text{sim}$ = 120\,ps& $t_\text{sim}$ = 5\,ns\\
 \hline

\multirow{7}{3cm}{L \\ $h$ = 30.0\,\AA} &  &  & $L_x$ = 36.89\,\AA\\
 &  &  & $L_y$ = 38.34\,\AA\\
 &  &  & $L_z$ = 90.65\,\AA \\
 &  &  & $N_\text{atoms}$ = 1080\\
 &  &  & $N_\text{water}$ = 1258\\
 &  &  & $t_\text{eq}$ = 2\,ns\\
 &  &  & $t_\text{sim}$ = 5\,ns\\
 \hline
 
 \multirow{7}{3cm}{XL \\ $h$ = 45.0\,\AA} &  &  & $L_x$ = 36.89\,\AA\\
 &  &  & $L_y$ = 38.34\,\AA\\
 &  &  & $L_z$ = 105.65\,\AA \\
 &  &  & $N_\text{atoms}$ = 1080\\
 &  &  & $N_\text{water}$ = 1968\\
 &  &  & $t_\text{eq}$ = 2\,ns\\
 &  &  & $t_\text{sim}$ = 5\,ns\\
 \hline

\end{tabular}
\end{table}

\clearpage

\section{Dielectric constant calculations}
\subsection{Subtracting the surface contribution}
In the AIMD simulations, the confining surfaces of the slits (graphene and hBN) contribute to the global polarization response due to the extension of their electronic clouds that partially overlaps with the one of water.
To eliminate this contribution, we corrected the calculated values by subtracting the local response of the system in the absence of water, that is, composed only of the confining surfaces (see snapshots in Fig. S1a-c), similarly as in the experiment \cite{fumagalli2018anomalously}.
First, we extracted the total local polarization response $\Delta P_{z}^{tot}$  from the total system.
We then extracted the polarization response of the confining surfaces $\Delta P_{z}^{CS}$ from a single calculation without water. 
This enabled us to obtain an approximation of the water response as $\Delta P_{z}^{H2O}=\Delta P_{z}^{tot}-\Delta P_{z}^{CS}$.
By combining it with the expression of the local dielectric constant, we get
\begin{equation}
\left[\epsilon^{H2O}_{\perp}(z)\right]^{-1} = \left[\epsilon^{tot}_{\perp}(z)\right]^{-1} + \left(1-\left[\epsilon^{CS}_{\perp}(z)\right]^{-1}\right).
\end{equation}
The result of this procedure is reported in Fig. S1 below, showing the dielectric constant profile obtained calculating a single configuration for the case of graphene slits (Fig. S1, profiles on the left) and hBN slits (Fig. S1, profiles on the right).

\begin{figure}[h!]
\centering
\includegraphics[scale=1.]{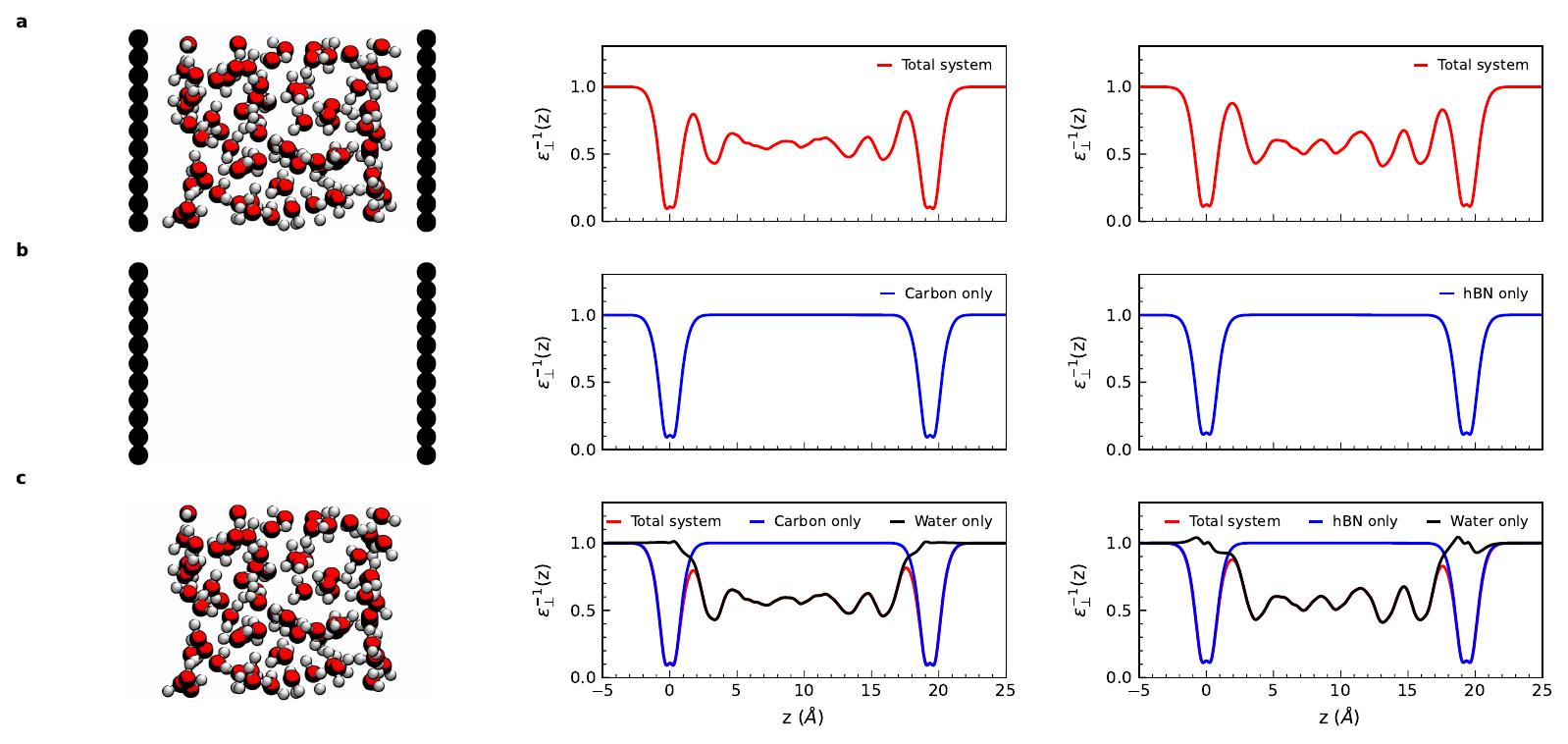}
\caption{\textbf{Calculation of water's dielectric constant profiles.} Snapshots (left) and corresponding dielectric profiles (right) for the total system (a), the system without water (b) and the profile (c) obtained after subtracting the polarization response of the graphene layer (left) and hBN layer (right) to the total system.}
\end{figure}

\subsection{Data convergence}
To check the equilibration of the simulations, we analyzed the time evolution of the dipole moment of the simulation boxes along with its running average, shown in Fig. S2a below for water inside a graphene slit for the M system.
The convergence of the total dipole moment was found to be very fast, as reported in a previous study on bulk water \cite{zhang_computing_2016}: equilibration and convergence is faster at finite D than at finite E.

\begin{figure}[h!]
\centering
\includegraphics[scale=0.7]{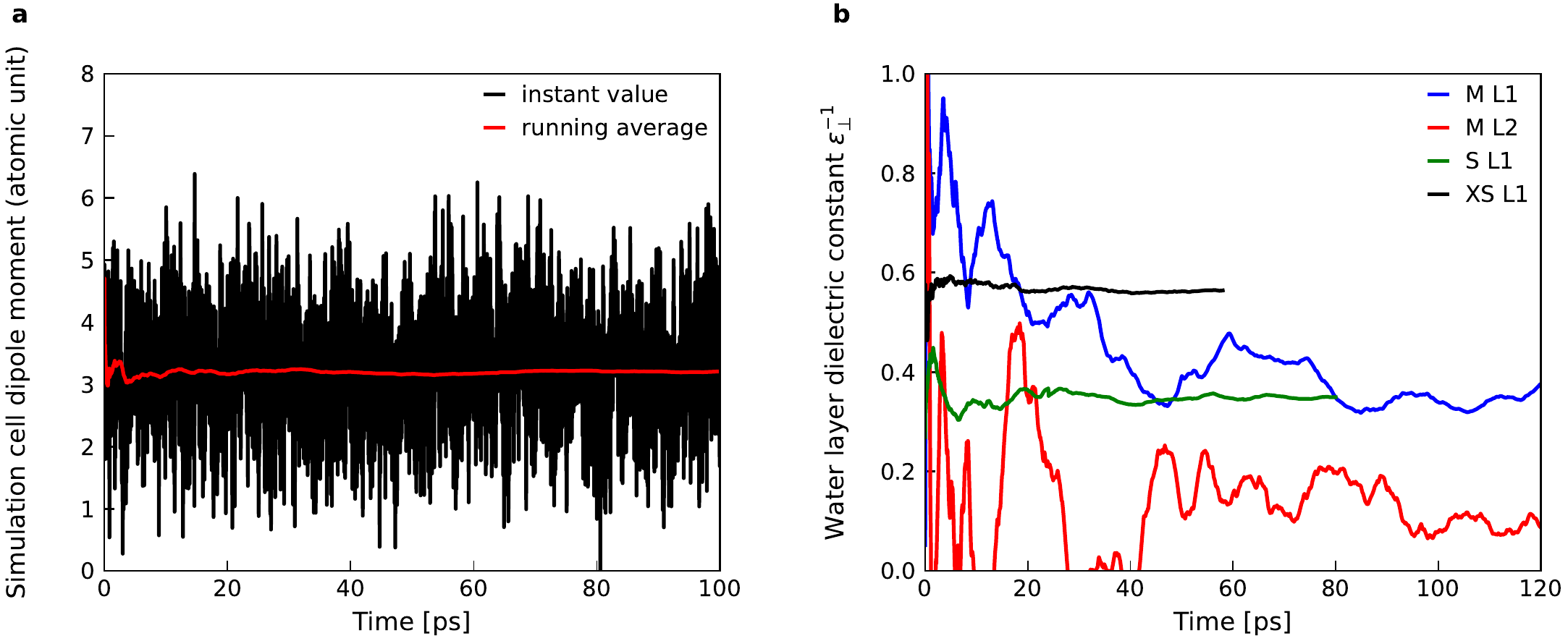}
\caption{\textbf{Convergence test of water's dielectric properties in AIMD simulations.} Convergence of the cell dipole moment calculated using the Berry phase for an electric displacement of -1.0 V.nm$^{-1}$ (a) with the corresponding running average of the total dipole moment for the M system and (b) running value of the dielectric constant of the interfacial water layers L1 and L2 for different water slab thicknesses.}
\end{figure}

We calculated the dielectric constant of the water layers reported in Fig. 1b in the main text from the running average of the dielectric constant, shown in Fig. S2b below, by taking the average of $\epsilon^{-1}_{\bot}$ in the long-time limit, corresponding to the last 20 ps of each simulation, except for the S system where the calculations were made on the last 10 ps. For the L system, we used the last 60 ps, as the response of the second layer requires more statistics to converge.
The error was estimated from the fluctuations of the running value of $\epsilon^{-1}_{\bot}$ around this average, and propagated to obtain the error in $\epsilon_{\bot}$.

We summarized the obtained dielectric constants of the first and second water layers (L1 and L2, respectively) for graphene and hBN slits in Table S2 below. For graphene, we also computed the dielectric constant $\epsilon_{\bot}^{C, PCD}$ by assuming that the charge distribution corresponds to the one of the SPCE model, i.e. a point charge distribution, in order to estimate the relative proportion of orientation and electronic effects.
We found that the obtained values of $\epsilon_{\bot}$ for graphene and hBN slits were equal within the error bar, except for the monolayer configuration which exhibits a slightly lower value for graphene. 
We obtained larger uncertainties in the dielectric constant for the second layer, L2, which we attribute to two factors. First, it arises from the higher error sensitivity of these calculations for higher values of $\epsilon_{\bot}$, that is, for a given uncertainty in $\epsilon^{-1}_{\bot}$, the uncertainty in $\epsilon_{\bot}$ increases with increasing $\epsilon_{\bot}$. 
Secondly, this may arise from the more dynamical character of the system with larger slit thickness, with more molecules exchanging between the layers that slow down the convergence.
The comparison with the dielectric constant obtained from the point charge distributions shows that most of the difference between the FFMD and AIMD calculations comes from the difference in the molecular orientation rather than electronic effects. 

To put these results in a broader perspective, we repeated the DFT simulations for the M graphene slit using the BLYP functionnal with the D3 correction. These calculations yielded a dielectric constant for the first interfacial water layer (L1) of $2.096 \pm 0.067$, a value in agreement with and even lower than what we found in this study. We note however that the response of the L2 layer did not converge even with 120 ps simulations.

\hfill \break
\begin{table}[h!]
\caption{\textbf{Dielectric constant of interfacial water layers.} Calculated dielectric constants of the interfacial water layers, L1 and L2, reported in Fig. 1b in the main text, for both graphene (first column) and hBN slits (second column) and various water slab thickness using AIMD simulations. The third column computes the dielectric constant from the AIMD configurations by assuming that the charge distribution corresponds to the one of the SPCE model, i.e. a point charge distribution. The error is obtained from the fluctuations of the running average in the long-time limit.
\label{tab:di-ele}}
\centering{}
\begin{tabular}{lc c c}
\hline
h (nm) & $\epsilon_{\bot}^{C}$ & $\epsilon_{\bot}^{hBN}$ & $\epsilon_{\bot}^{C, PCD}$\\ \hline
\multicolumn{4}{c}{L1} \\
1.935 & 2.94 $\pm$ 0.12 & 2.80 $\pm$ 0.12 & 2.80 $\pm$ 0.33\\
0.911 & 2.874 $\pm$ 0.013 & 2.871 $\pm$ 0.015 & 3.38 $\pm$ 0.16\\
0.663 & 1.7805 $\pm$ 0.0054  & 2.344 $\pm$ 0.010 & 2.116 $\pm$ 0.039\\
\multicolumn{4}{c}{L2} \\
1.935 & 10.55 $\pm$ 1.62 & 11.07 $\pm$ 8.03 & 7.80 $\pm$ 2.76\\
\hline
\end{tabular}
\end{table}

\newpage

\subsection{Impact of the dividing surface}
The dividing surface, that is, the domain boundary over which the dielectric profile is integrated, is not well-defined on the atomic scale due to the non-continuous nature of the interface between the confining layer and the water phase. We investigated the effect of the choice of the dividing surface by considering four positions, indicated in Fig. S3a (dashed lines): the confining surface (CS), the one used in the main text corresponding to the atomic position of the confining layer (e.g. carbon atoms for graphene); the electronic density crossover (EC), corresponding to the local density maximum between the carbon wall and the water phase; the full density (FD) surface, corresponding to the symmetric position of the EC surface with respect to the confining surface; and the jellium edge (JE) \cite{smith1989distance}, corresponding to the extension of the electronic cloud of the confining material in a mean-field approach. 

Figure S3b shows the calculated inverse dielectric constant for each position of the dividing surface against the experimental data as a function of the water slab thickness. The calculated values are sensitive to the dividing surface position, and this is due to the change in the domain volume, as the dipole moment remains largely unchanged. In particular, shifting the dividing surface from the atomic position towards the vacuum, as in the case of the FD surface, increases the volume at constant dipole moment, thus decreasing the resulting polarization, which in turn leads to a decrease in the obtained $\epsilon_{\bot}$. The opposite occurs when the dividing surface is shifted towards the liquid phase, as in the case of EC and JE surfaces: the polarization is increased with decreasing the volume, and in turn $\epsilon_{\bot}$ increases up to infinity and even becomes negative. In this study we show that our calculations yield values of $\epsilon_{\bot}$ in good agreement with the experimental values by simply taking the dividing surface at its physical position - the atomic position of the confining layer (CS) - and subtracting the electronic cloud contribution beyond the surface, as explained in section S2.A above, without any need of shifting the dividing surface to a different position. We also note that for all the dividing surfaces considered here, our AIMD calculations yielded smaller $\epsilon_{\bot}$ than the values obtained from force-field simulations.

\begin{figure}[h!]
\centering
\includegraphics[scale=0.79]{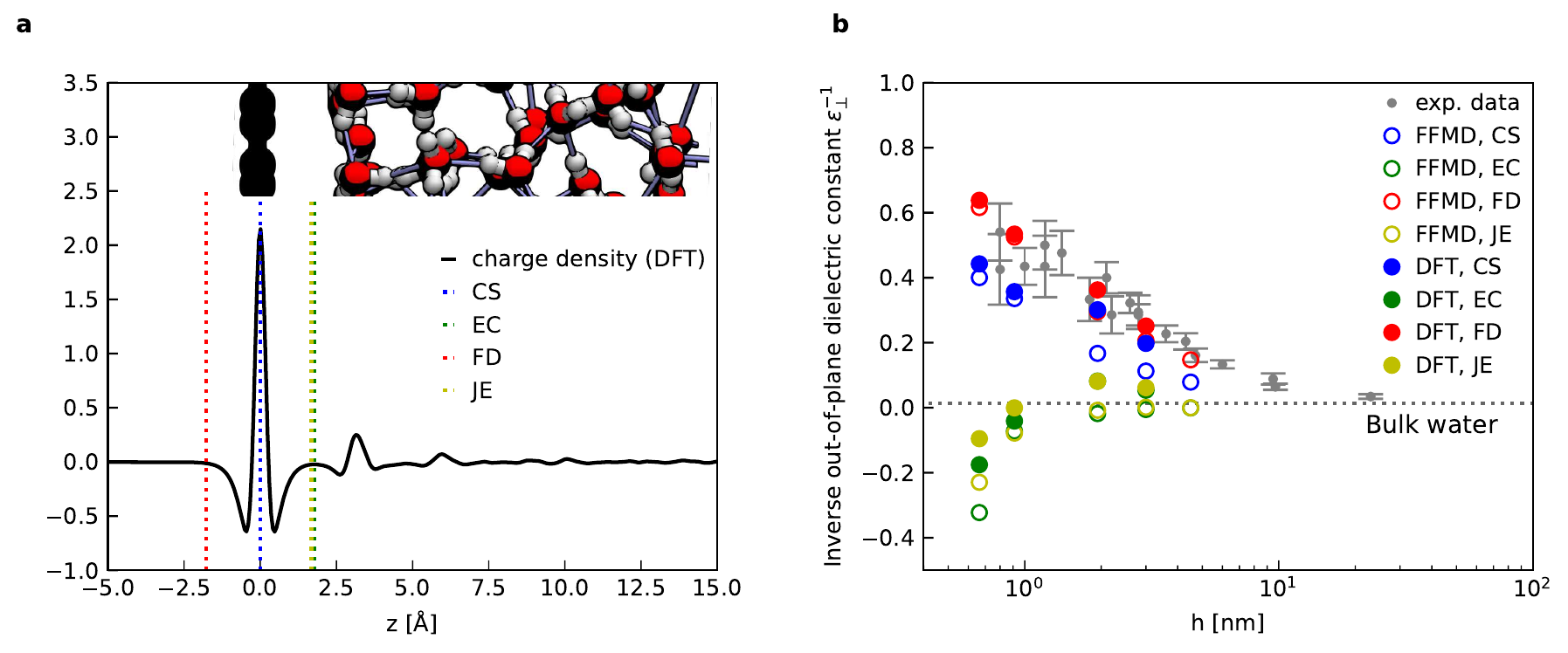}
\caption{\textbf{Impact of the dividing surface.} (a) Examples of positions of the dividing surface (dashed vertical lines) at the carbon atom side and calculated charge density profile using DFT calculations (solid line). A snapshot of the system is provided as a guidance. (b) Calculated inverse dielectric constant of water for various dividing surfaces shown in (a) using DFT and FFMD calculations.
\label{fig:div-sur}}
\end{figure}

\clearpage

\section{Additional force-field simulations}
\begin{figure}[h]
\centering
\includegraphics[scale=0.7]{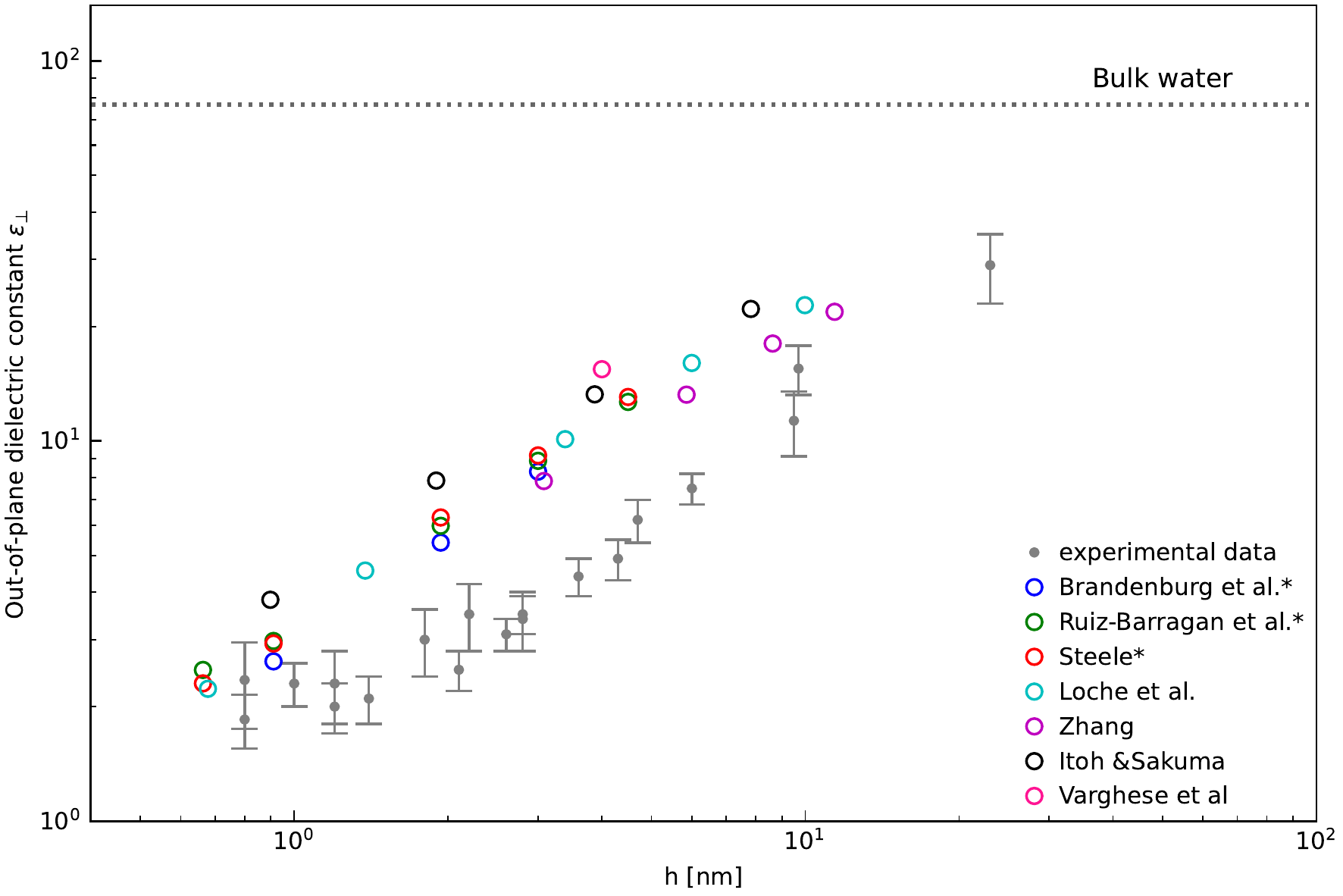}
\caption{\textbf{Impact of the force-field choice.} Calculated out-of-plane dielectric constant of water confined in graphene slits using different force-fields in literature, reported in the following references: Brandenburg \textit{et al.} \cite{brandenburg2019interaction}, Ruiz-Barragan \textit{et al.} \cite{ruiz2020quantifying}, Steele \cite{steele1974interaction}, Loche \textit{et al.} \cite{loche2020universal}, Zhang \cite{zhang2018note}, Itoh \& Sakuma \cite{itoh2015dielectric}, Varghese \textit{et al.} \cite{varghese2019effect}. Symbols labeled with a star (*) correspond to simulations made in this study, while the rest of simulated data are reported here from the references.  Grey symbols are the experimental data reported here from Ref.\cite{fumagalli2018anomalously}.
\label{fig:ff-res}}
\end{figure}

We compared the out-of-plane dielectric constant, $\epsilon_{\bot}$, of confined water calculated from our AIMD simulations with values obtained from FFMD simulations with the aim of evaluating the performance of our simulations and the impact of different force-fields. Figure~\ref{fig:ff-res} shows different FFMD simulations, either reported in the literature or carried out in this study, of water confined in graphene slits as a function of the water slab thickness against the experimental data of Ref.\cite{fumagalli2018anomalously}.  
The water models used in the FFMD simulations reported here from the literature are SPCE \cite{berendsen1987missing} (used in \cite{zhang2018note,varghese2019effect,loche2020universal}) and KKY \cite{kumagai1994interatomic} (used in \cite{itoh2015dielectric}), a more complex model that allows intramolecular vibrations. Graphite and graphene are modeled by rigid carbon walls using the carbon-water two body interactions.
The force-field models are parametrized from various sources, namely, DFT calculations \cite{itoh2015dielectric} using a Buckingham potential based on \cite{cortes2013interaction}, properties of water droplets on graphene \cite{varghese2019effect} using a carbon-oxygen van der Waals potential \cite{werder_watercarbon_2003}, or properties of biomolecules \cite{loche2020universal} using the GROMOS53A6 force field \cite{oostenbrink2004biomolecular} (a carbon-oxygen Lennard-Jones interaction) or a Lennard-Jones potential using varying parameters in \cite{zhang2018note}.
In the FFMD simulations carried out in this study, we used three different water-carbon interaction potentials, with two 2-body potentials, one adjusted on DFT which was used for the data reported in the main text \cite{ruiz2020quantifying} and one on \cite{steele1974interaction}, and a carbon wall potential adjusted on water adsorption from quantum Monte Carlo data \cite{brandenburg2019physisorption}. 
Figure \ref{fig:ff-res} clearly shows that our simulations gives similar results as reported in literature, with all FFMD simulations yielding a similar decrease of $\epsilon_{\bot}$ as a function of the water slab thickness, irrespective of the force field used. In particular, they all match for water under strong confinement, when the distance between the confining surface is less than 1 nm. This indicates that when the strength of the confinement becomes large, the exact details of the interaction have no impact. Indeed, under strong confinement, the orientation of the water molecule is constrained in the in-plane direction due to the small space available in the slit, thus the  ability of the water dipole to reorient with the external electric field is strongly suppressed. For larger water slabs, the shift in the predicted $\epsilon_{\bot}$ associated with different force fields remains substantially negligible.    

\hfill \break

\begin{figure}[h!]
\centering
\includegraphics[scale=0.7]{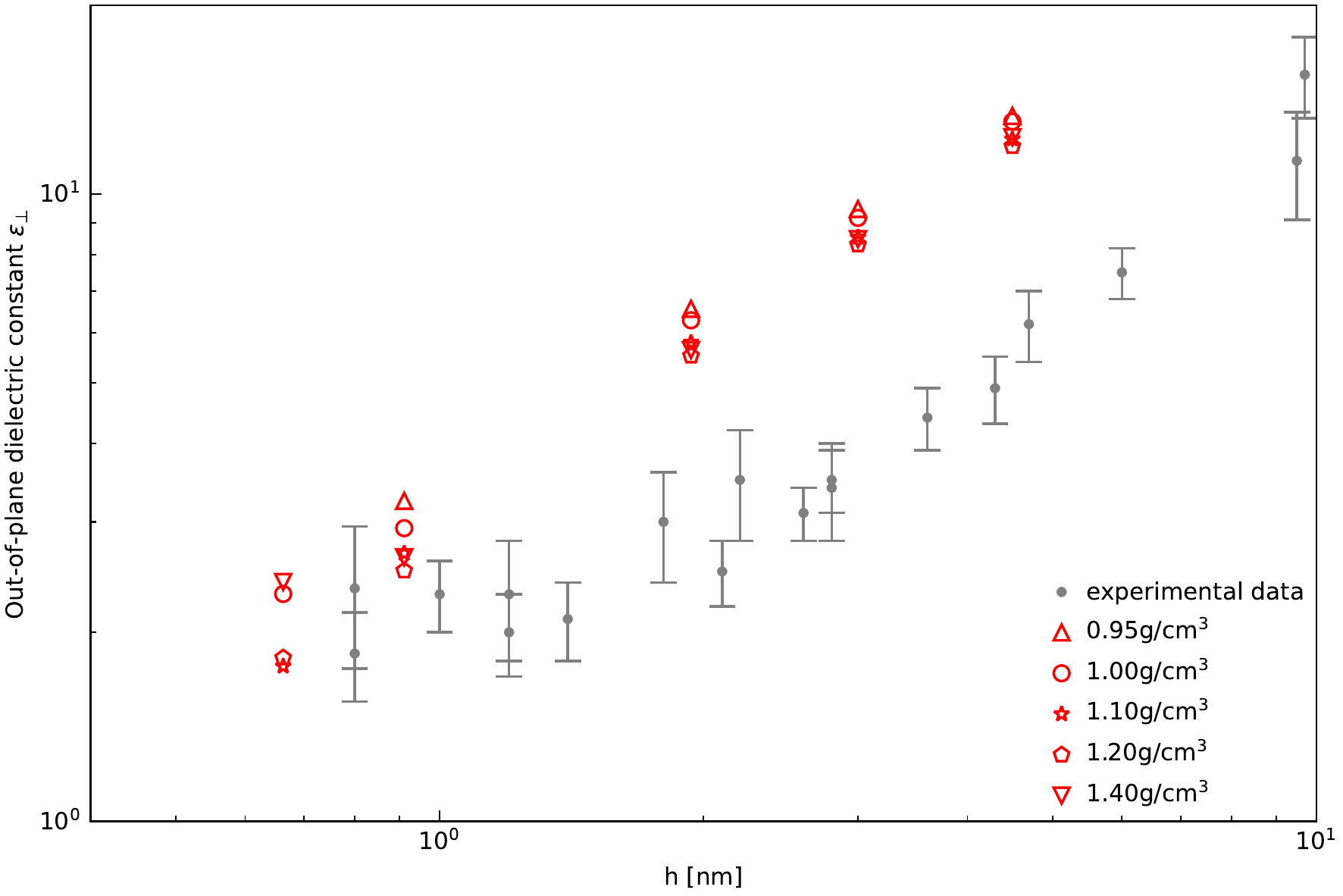}
\caption{\textbf{Impact of bulk water density on the dielectric constant of confined water.} Calculated dielectric constant of water confined in graphene slits for different values of bulk water density. Grey symbols are the experimental data reported here from Ref.\cite{fumagalli2018anomalously}.
\label{fig:density-effect}}
\end{figure}

We also investigated the effect of the water density on the calculated dielectric  constant of the water confined inside graphene slits, as the water density in the slits used in the experiment is not known. 
To this aim, we run FFMD simulations with different values of water density ranging from 0.95 to 1.4 g/cm$^{3}$, corresponding to lateral pressures between -0.3 kbar and around 6 kbar inside the confined region. Results are plotted in Fig.~\ref{fig:density-effect}, again against the experimental data, clearly showing that the change in water density range considered here has only a minor impact on the predicted values of $\epsilon_{\bot}$ of confined water.  
The small variation in $\epsilon_{\bot}$ observed here arises from the competition between two effects. On the one hand, increasing the water density increases the number of molecular dipoles per volume unit. If the response of individual dipole moments remains unchanged, the polarization response is then expected to increase. On the other, increasing the water density would increase the rigidity of the hydrogen bonding network, which in turn decreases the ability of the water dipoles to reorient with the external electric field.
Our simulations show a small decrease in $\epsilon_{\bot}$ with increasing the water density up to 1.1 g/cm$^{3}$, suggesting the increased connectivity of the hydrogen bonding network is dominant at relatively low density values. For density larger than 1.1 g/cm$^{3}$, the trend reverses, indicating a small increase in $\epsilon_{\bot}$, thus the increase of dipole density becomes the dominant effect. 
However, the variations obtained here clearly show that both effects are not significant, with a decrease in $\epsilon_{\bot}$ of only 10\% for a change in water density of 20\%. 

\clearpage

\section{Hydrogen bond orientation}
For the analysis of the hydrogen bonding network inside the slit described in the main text, we used the hydrogen bond definition based on a distance criterion between the two oxygen atoms $r_\text{OO}$ and an angle criterion on the $\theta_{\text{HO}_{d}\text{O}_{a}}$ angle, where O$_{d}$ is the oxygen of the donor molecule and O$_{a}$ is the oxygen of the acceptor \cite{Hbond_criteria}.
The two molecules form a hydrogen bond if $r_\text{OO} \leq 3.5$ \AA, and $\theta_{\text{HO}_{d}\text{O}_{a}} \leq 30^{\circ}$.
The direction of the hydrogen bond is then defined as the direction of the $\overrightarrow{\text{O}_{d}\text{O}_{a}}$  vector.

\begin{figure}[h!]
\centering
\includegraphics[scale=0.85]{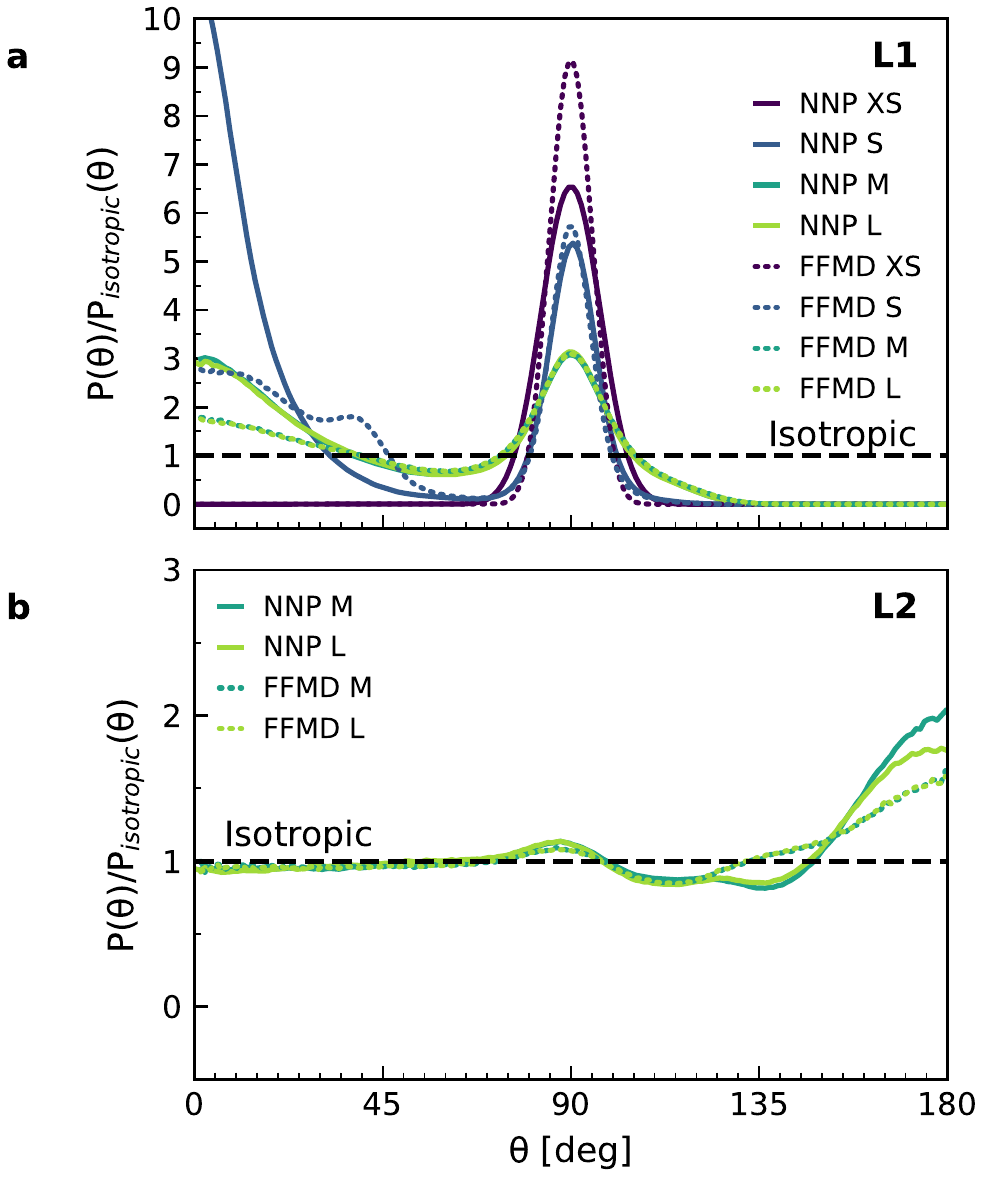}
\caption{\textbf{Orientation of hydrogen bonds in the interfacial water.} Orientation distribution with respect to the normal direction of the hydrogen bonds for water (a) in the first and (b) the second interfacial layer (L1 and L2) near the graphene surface, normalized by the isotropic distribution, calculated using NNP (solid lines) and FFMD (dashed lines) simulations for various slab thicknesses. The direction corresponds to the donor oxygen-acceptor oxygen.
\label{fig:si-hbond-ori}} 
\end{figure}

Figure~\ref{fig:si-hbond-ori} compares the H-bond distributions obtained from NNP simulations, reported in Fig. 3 in the main text, with the distributions obtained from our FFMD simulations in the first two interfacial water layers near the graphene interface. From the comparison between the two data set, it can be seen that the obtained orientation is similar, but the NNP simulations yield a larger proportion of out-of-plane hydrogen bonds and in a smaller angular window in the first interfacial layer (L1). This indicates that a more rigid hydrogen-bonding network is predicted by NNP simulations as opposed to FFMD simulations. 

\hfill \break

\clearpage

\clearpage

\bibliographystyle{naturemag}
\bibliography{NewLib}